\documentclass[journal]{IEEEtran}
\ifCLASSINFOpdf
\else
\fi
\usepackage{color}
\usepackage{graphicx}
\usepackage{amsmath}
\usepackage{tabularx}
\usepackage{multirow}
\usepackage{booktabs}
\usepackage{cite}
\usepackage{datetime}
\usepackage{changepage}

\usepackage{lipsum}
\usepackage{mathtools}

\usepackage{cuted}

\newcommand{\sci}[3][10]{{#2}$\times${#1}${}^\mathrm{#3}$}

\newcommand{\md}{\mathcal{D}}
\newcommand{\diff}[2]{\frac{\mathrm{d}{#1}}{\mathrm{d}{#2}}}

\newcommand{\x}{\hspace{2.5mm}}
\newcommand{\xxdef}{\stackrel{\text{def}}{=}}

\newcommand{\rcolor}[1]{\textcolor{black}{#1}}

\begin{document}
	%
	\title{Hybrid QSS and Dynamic Extended-Term Simulation Based on Holomorphic Embedding}
	%
	%
	%
	
	\author{Rui~Yao,~\IEEEmembership{Senior~Member,~IEEE,}~and~Feng~Qiu,~\IEEEmembership{Senior~Member,~IEEE}
		\thanks{This work was supported by the Advanced Grid Modeling (AGM) program of U.S. Department of Energy.}
		\thanks{R. Yao and F. Qiu are with the Division of Energy Systems, Argonne National Laboratory, Lemont 60439, USA. (emails: ryao@anl.gov, fqiu@anl.gov).}
	}
	
	%
	%

	\markboth{
	}%
	{Shell \MakeLowercase{\textit{et al.}}: ZZZZZZ}
	%



	\maketitle
	
	\begin{abstract}
		Power system simulations that extend over a time period of minutes, hours, or even longer are called \textit{extended-term} simulations. As power systems evolve into complex systems with increasing interdependencies and richer dynamic behaviors across a wide range of timescales, extended-term simulation is needed for many power system analysis tasks (e.g., resilience analysis, renewable energy integration, cascading failures), and there is an urgent need for efficient and robust extended-term simulation approaches. The conventional approaches are insufficient for dealing with the extended-term simulation of multi-timescale processes. This paper proposes an extended-term simulation approach based on the holomorphic embedding (HE) methodology. Its accuracy and computational efficiency are backed by HE's high accuracy in event-driven simulation, larger and adaptive time steps, and flexible switching between full-dynamic and quasi-steady-state (QSS) models. We used this proposed extended-term simulation approach to evaluate bulk power system restoration plans, and it demonstrates satisfactory accuracy and efficiency in this complex simulation task.
	\end{abstract}
	

	%
	\IEEEpeerreviewmaketitle

	\section{Introduction}
	
	\IEEEPARstart{P}{ower} systems have various kinds of networked components as well as complex behaviors. The power system dynamics have multiple distinct time scales\cite{chen2008variable,7254205}. For example, the timescales of the fast transients can be less than 0.01 s, and the actions of system-wide control (e.g., automatic generation control or AGC)\cite{ma2013simagc} and some mechanical and thermal-driven processes \cite{chen2017numerical} are usually in the timescales of seconds to several minutes. Changes in load levels and the economic dispatch actions take minutes to hours, and due to the interdependencies among the system components and various external impacts (e.g., weather, vegetation, natural disasters), complex event chains may occur. These complexities call for panoramic simulations of complex event processes involving various disturbances, system responses and control measures, and traditional security analysis methods and tools are insufficient for such computation tasks. For example, the transient stability analysis only assumes a single fault, and the duration is usually within 1 minute, which ignores the longer-timescale dynamics \cite{korkali2017gmlc} such as AGC and dispatch. On the other hand, steady-state security analysis based on power flow models for longer-timescale analysis cannot capture system dynamics\cite{vournas1998modelling}. Because many security concerns in the power systems involve complex and extended-term processes\cite{fu2011high}, it is imperative to find new methods for robust and efficient extended-term simulation.
	
	Traditional numerical computation methods have major limitations for extended-term simulations. The traditional numerical integration methods for solving differential equations are lower-order methods\cite{yao2019efficient}, and their efficiency is limited by tiny time steps. Such methods cannot flexibly adapt to the variations of dominant timescales. Moreover, the Newton-Raphson approach is commonly used to solve nonlinear equations, but its convergence highly depends on the initial solution and frequently faces non-convergence issues when solving large deviations of system states. The \textit{Holomorphic embedding} (HE) is an emerging approach \cite{rao2016holomorphic,liu2017online,yao2019efficient} for extended-term simulation. HE has shown promising performance in the steady-state \cite{rao2016holomorphic,liu2017online} and dynamic analysis \cite{yao2019efficient} tasks of power systems. The HE adopts an analytical approximate solution as a continuous function in the time domain, which provides a very flexible selection of time steps, and HE guarantees convergence when solving nonlinear equations, which avoids the computation failures in extended-term simulation. 
	
	More important, HE shows natural advantages for handling events and multi-timescale simulation because of its analytical form in the time domain. This paper will show the promising potential of HE for hybrid extended-term simulation based on a simulation framework combining steady-state and dynamic simulation. The dynamic simulation can be performed during system transients, while quasi-steady-state (QSS) modeling can be adopted after the transients fade away. Switching from dynamic to QSS simulation can be efficiently performed with HE by using the HE solution parameters, which avoids the extra simulation burden in traditional approaches based on evaluating the trajectory variations. The extended-term simulation can be utilized for complex analysis tasks such as resilience analysis\cite{wang2016resilience,panteli2015grid, huang2017integration}, cascading outages\cite{7254205,fu2011high,song2015dynamic}, restoration \cite{qiu2017integrated}, and renewable energy control\cite{qureshi2019fast}.
	
	The contributions of this paper are threefold:
	
	(1) We present the HE formulations for the simulation of atomic events in power system analysis. The atomic event simulators constitute the extended-term simulation.
	
	(2) We propose a hybrid simulation scheme that switches between dynamic simulation and QSS simulation. Switching from dynamic simulation to QSS simulation can be directly determined from the analytical form of the HE solution and thus is much more efficient than the traditional practices.
	
	(3) We propose an extended-term event-driven simulation framework based on holomorphic embedding. Thanks to the analytical nature of HE solutions, the event scheduler can handle various types of events with enhanced accuracy.
	
	The rest of the paper is organized as follows: Section II elucidates how to simulate some typical types of events (atomic events) using HE; Section III presents a hybrid simulation framework combining the dynamic and steady-state simulation with HE; Section IV is the overall procedure of the event-driven extended-term simulation; Section V is the case studies; and Section VI is the conclusion of the paper.
	
	\section{Atomic Event Simulator Based on HE}
	The extended-term simulation comprises several kinds of basic events, such as adding/cutting elements and ramping events. Here we call them atomic events. This section will start with a fundamental formulation of HE, and will then introduce the methods for simulating atomic events in HE.
	\subsection{Brief overview of HE for power system analysis}
	A power system can be modeled by differential algebraic equations (DAEs) in segments in the time domain:
	\begin{equation}\label{eqn:general_dae}
	\begin{aligned}
	\dot{\mathbf{x}}&=\mathbf{f}(\mathbf{x}, \mathbf{y}, \mathbf{p})\\
	\mathbf{0}&=\mathbf{g}(\mathbf{x}, \mathbf{y}, \mathbf{p})
	\end{aligned}
	\end{equation}
	where $\mathbf{x}$ is the state variable, $\mathbf{y}$ is the algebraic variable and $\mathbf{p}$ represents the system parameter and control variable. $\mathbf{p}$
	may change with time, and within a time interval, $\mathbf{p}(t)$ can be represented by or approximated as a power series of time:
	\begin{equation}\label{eqn:ramping_variables}
	\mathbf{p}(t)=\sum_{k=0}^{N}\mathbf{p}[k]t^k.
	\end{equation}
	Eq. (\ref{eqn:general_dae}) can be solved by using HE \cite{yao2019efficient}. The HE solution has the following power series form:
	\begin{equation}\label{eqn:he_sol}
	\mathbf{x}(t)\approx\sum_{k=0}^{N}\mathbf{x}[k]t^k,~
	\mathbf{y}(t)\approx\sum_{k=0}^{N}\mathbf{y}[k]t^k
	\end{equation}
	or its corresponding Pad\'e approximation \cite{yao2019vectorized}. Therefore, within a segment of time domain, the system dynamics are approximated as a continuous function of time, showing that HE is completely different from the traditional numerical integration methods on discrete time points.
	The HE formulation of a system with loads and generators can be written as:
	\begin{equation}\label{eqn:power_flow_eqn_nt_pvpq}
	\resizebox{0.9\hsize}{!}{$\displaystyle{(P_i(t)-jQ_i(t))W_i^*(t)-\sum_{l}Y_{il}V_l(t)-I_{Li}(t)+I_{Gi}(t)=0,}$}
	\end{equation}
	where $V_i$ is the voltage phasor on bus $i$, whose reciprocal is $W_i$. $Y_{il}$ is the element of row $i$ and column $l$ in the admittance matrix $\mathbf{Y}$. The constant-PQ loads are $P_i$ and $Q_i$ terms (positive values denote generation and negative values denote loads), and the constant-impedance loads are merged into $\mathbf{Y}$. The current of other loads (e.g., constant-current loads, induction motor loads) on bus $i$ is represented by $I_{Li}$, and the current of all the synchronous machines is represented by $I_{Gi}$.
	
	The events in the power system can be classified into several kinds of ``atomic'' events, such as ramping events, adding elements, cutting elements, and instantly changing element parameters. HE-based modeling and simulation of the atomic events will be elaborated upon in the following sub-sections.
	
	\subsection{Simulation of system dynamics with ramping events}
	The continuous ramping of control variables or system configurations, such as the ramping of generator outputs, AVR reference signals, or loads, can be represented with polynomials of time. 
	For example, in the turbine governor, 
	the reference mechanical torque $T_{Mi0}$ can be adjusted by the external control (e.g., automatic generation control or the ramping command from operators). According to (\ref{eqn:ramping_variables}), $T_{Mi0}(t)$ can be expressed or approximated as a polynomial of time $t$:
	\begin{equation}\label{eqn:tg_tmech0_t}
	T_{Mi0}(t)=\sum_{k=0}^{N_{tg}}T_{Mi0}[k]t^k,
	\end{equation}
	where $T_{Mi0}[k]$ are known coefficients. These coefficients are used directly to derive the equations of HE coefficients for the unknown variables.
	
	\subsection{Simulating instant-switching events}
	\subsubsection{General principles}
	An instant switch here means an event of instantly adding or tripping components, corresponding to switch opening/closing operations. Switching instantly changes algebraic variables (e.g., bus voltages), and creates momentum for state variables to change. Therefore, transients usually follow the instant-switching events, and on the occurrence of an instant-switching event, the dynamic model should be used for a period of time until the system reaches a new steady state. 	
	The basic idea of solving instant switches using HE is to construct and solve HE formulations so that $\alpha=0$ corresponds to the pre-switch instant and $\alpha=1$ corresponds to the post-switch state. The next subsections will deal with some typical types of events in detail.

	\subsubsection{Adding elements}
	An added element could be a single device with one terminal (e.g., a generator, static load, induction motor) or multiple terminals (e.g., a transmission line), or even more broadly, a subsystem. The model of an added element may have its own state variables and algebraic variables, and thus its own DAEs. A typical example is a synchronous generator and its corresponding DAEs. After the instant of adding the element, the original system and the added element need to satisfy the boundary conditions of voltage and current at the connection points. Generally, we assume that the state variables of the new element are $\mathbf{x}_E$, and the internal algebraic variables are $\mathbf{y}_E$, the terminal voltage of the element is $\mathbf{V}_E$, and the current is $\mathbf{I}_E$. The behavior of the element can be modeled as:
	\begin{equation}\label{eqn:general_one_terminal_element}
	\begin{aligned}
	\dot{\mathbf{x}}_E&=\mathbf{f}_E(\mathbf{x}_E, \mathbf{y}_E, \mathbf{V}_E,\mathbf{I}_E,\mathbf{p}_E)\\
	\mathbf{0}&=\mathbf{g}_E(\mathbf{x}_E, \mathbf{y}_E,\mathbf{V}_E,\mathbf{I}_E,\mathbf{p}_E)
	\end{aligned}
	\end{equation}
	
\begin{figure}[htb]
		\centering
		\includegraphics[clip=true,scale=0.14]{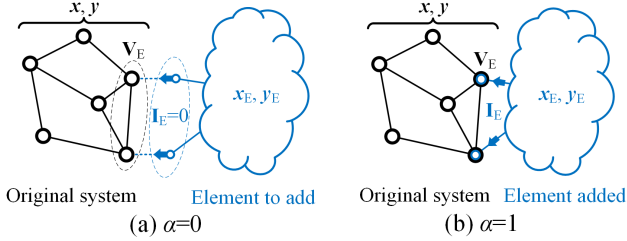}\\
		\caption{HE solving post-switch state.}
		\label{fig:add_element}
	\end{figure}

	At the instant after switching, the post-switching states of the system should satisfy the algebraic equations of the original network and the added element:
	
	\begin{equation}\label{eqn:general_one_terminal_element_sys}
	\begin{aligned}
	\mathbf{0}&=\mathbf{g}(\mathbf{x}, \mathbf{y},\mathbf{V}_E,\mathbf{I}_E,\mathbf{p})\\
	\mathbf{0}&=\mathbf{g}_E(\mathbf{x}_E, \mathbf{y}_E,\mathbf{V}_E,\mathbf{I}_E,\mathbf{p}_E)
	\end{aligned}
	\end{equation}
	where $\mathbf{g}$ represents the algebraic equations of the original system.  
	According to the implicit function theorem, normally the current of the element $I_E$ can be explicitly written as
	\begin{equation}\label{eqn:general_one_terminal_element_exp}
	\mathbf{I}_E=\mathbf{g}_{EI}(\mathbf{x}_E, \mathbf{y}_E, \mathbf{V}_E,\mathbf{p}_E)
	\end{equation}
	To solve the post-switch state, we build the HE formulation: 
	\begin{equation}\label{eqn:one_terminal_solve_he}
	\begin{aligned}
	\mathbf{I}_E(\alpha)&=\mathbf{g}_{EI}(\mathbf{x}_E(\alpha), \mathbf{y}_E(\alpha),\mathbf{V}_E(\alpha),\mathbf{p}_E(\alpha))\\
	\mathbf{0}&=\mathbf{g}(\mathbf{x}, \mathbf{y}(\alpha),\mathbf{V}_E(\alpha),\mathbf{I}_E(\alpha),\mathbf{p})
	\end{aligned}
	\end{equation}
	where, as Fig. \ref{fig:add_element} shows, $\alpha=0$ corresponds to the pre-switch state and requires that $\mathbf{I}_E(\alpha)=\mathbf{0}$. $\alpha=1$ corresponds to the post-switch state. Thus, solving (\ref{eqn:one_terminal_solve_he}) at $\alpha=1$ will derive the state after adding the element. 
	The solutions in (\ref{eqn:one_terminal_solve_he}) apply to various elements. An example will illustrate the idea and the method. Many elements can be modeled as Thevenin models:
	\begin{equation}\label{eqn:one_terminal_thevenin} 
	\left[\begin{array}{c}
	I_{xi}\\I_{yi}
	\end{array}\right]
	=
	\left[
	\begin{array}{cc}
	Y_{i11}  & Y_{i12}\\
	Y_{i21}  & Y_{i22}\\
	\end{array}
	\right]
	\left[\begin{array}{c}
	E_{xi}-V_{xi}\\E_{xi}-V_{yi}
	\end{array}\right]
	\end{equation}
	where $V_{xi}$ and $V_{yi}$ are the real and imaginary parts of the terminal voltage, and $I_{xi}$ and $I_{yi}$ are the real and imaginary parts of the terminal current. For instance, a synchronous generator has the following form \cite{yao2019efficient}:
	\begin{equation}\label{eqn:sg_generalize_transform}
	\left[\begin{array}{c}
	I_{xi}\\I_{yi}
	\end{array}\right]=
	\mathbf{M}(\delta_i)
	\mathbf{Y}_{gi}^{-1}
	\mathbf{M}(\delta_i)^{\mathrm{T}}
	\left(
	\mathbf{M}(\delta_i)
	\left[
	\begin{array}{c}
	\varepsilon_{di} \\
	\varepsilon_{qi} \\
	\end{array}
	\right]-
	\left[
	\begin{array}{c}
	V_{xi} \\
	V_{yi} \\
	\end{array}
	\right]
	\right)
	\end{equation}
	where $\varepsilon_{di}$ and $\varepsilon_{qi}$ are the internal potentials on d- and q- axes, $\delta_{i}$ is the rotor angle, and
	\begin{equation}\label{eqn:sg_parameters}
	\mathbf{M}(\delta_i)=\left[
	\begin{array}{cc}
	\sin\delta_i  & \cos\delta_i\\
	-\cos\delta_i & \sin\delta_i\\
	\end{array}
	\right],
	\mathbf{Y}_{gi}=
	\left[
	\begin{array}{cc}
	\rho_{si} & -\chi_{qi}\\
	\chi_{di} & \rho_{si}\\
	\end{array}
	\right]
	\end{equation} 
	where $\rho_{si}$ is the internal equivalent resistance of the generator, and $\chi_{di}$ and $\chi_{qi}$ are the internal equivalent impedance on the d- and q- axes, respectively.
	For the Thevenin model (\ref{eqn:one_terminal_thevenin}), the HE formulation for solving the post-switch state is:
	\begin{equation}\label{eqn:one_terminal_thevenin_he}
	\left[\begin{array}{c}
	I_{xi}(\alpha)\\I_{yi}(\alpha)
	\end{array}\right]
	= 
	\alpha\cdot
	\left[
	\begin{array}{cc}
	Y_{i11}  & Y_{i12}\\
	Y_{i21}  & Y_{i22}\\
	\end{array}
	\right]
	\left[\begin{array}{c}
	E_{xi}-V_{xi}(\alpha)\\E_{xi}-V_{yi}(\alpha)
	\end{array}\right]
	\end{equation}
	and the solution of (\ref{eqn:one_terminal_thevenin_he}) at $\alpha=1$ is the post-switch state.

	\subsubsection{Cutting elements}
	Cutting elements is generally the inverse process of adding the elements. As in (\ref{eqn:general_one_terminal_element_sys}) and (\ref{eqn:general_one_terminal_element_exp}) the system with the element to be cut follows the equation:
	\begin{equation}\label{eqn:mult_terminal_element_cut}
	\begin{aligned}
	\mathbf{0}&=\mathbf{g}(\mathbf{x}, \mathbf{y},\mathbf{V}_E,\mathbf{I}_E,\mathbf{p})\\
	\mathbf{I}_E&=\mathbf{g}_{EI}(\mathbf{x}_E, \mathbf{y}_E, \mathbf{V}_E,\mathbf{p}_E),
	\end{aligned}
	\end{equation}
	where the internal state variables and algebraic variables to be cut are denoted as $\mathbf{x}_E$ and $\mathbf{y}_E$, and the state variables and algebraic variables of the rest of the system are $\mathbf{x}$ and $\mathbf{y}$. The boundary voltage is $\mathbf{V}_E$, which is also included in $\mathbf{y}$. However, unlike the added elements, the cut element will no longer be of interest (i.e., it is regarded as unrecoverable), so a simple HE formulation based on equivalent admittance can be constructed. On boundary buses, the voltage and current at the pre-switch state are $\mathbf{V}_E$ and $\mathbf{I}_E$, respectively, which is equivalent to a set of shunt admittances $\mathbf{Y_{E}}=\md(I_{Ei}/V_{Ei})$. Thus (\ref{eqn:mult_terminal_element_cut}) can be replaced with
	\begin{equation}\label{eqn:mult_terminal_element_cut_simp}
	\mathbf{I}_E=\mathbf{Y_{E}}\mathbf{V}_E
	\end{equation}
	and the according to (\ref{eqn:mult_terminal_element_cut}), the HE formulation for solving the post-switch state of the rest of the system is:
	\begin{equation}\label{eqn:mult_terminal_cut_solve_he_simp}
	\mathbf{0}=\mathbf{g}(\mathbf{x}, \mathbf{y}(\alpha), \mathbf{V}_E(\alpha),(1-\alpha) \mathbf{Y_{E}}\mathbf{V}_E(\alpha),\mathbf{p}).
	\end{equation}
	
	The formulation (\ref{eqn:mult_terminal_cut_solve_he_simp}) can reduce the computational burden because the internal states of the cut elements are omitted.
	
	\subsubsection{Changing element parameters}
	Some instant switches involve a change of element parameters $\mathbf{p}$, usually because of sudden changes inside elements. A typical example is a fault, which changes the admittance parameters and thus changes the admittance matrix\cite{yao2019efficient}. To solve the post-switch state, assume the parameters change from $\mathbf{p}$ to $\mathbf{p}'$. The following HE formulation can be constructed and solved:
	\begin{equation}\label{eqn:parameter_change}
	\mathbf{0}=\mathbf{g}(\mathbf{x}, \mathbf{y}(\alpha),\alpha\mathbf{p}'+(1-\alpha)\mathbf{p})
	\end{equation}
	
	\section{Steady-state \& Dynamic Hybrid Simulation}
	\subsection{Switching from dynamic to quasi-steady-state (QSS) models}
	To change dynamic models to steady-state models, the prerequisite is that the system be approximately in a steady state (which will be addressed in Section \ref{subsec:he_switching}). The elements need to be noted are the synchronous generators. Usually, the generators are equipped with automatic voltage regulators (AVRs) to maintain terminal voltage, so they can be converted to PV buses in the QSS model.
	
	The QSS model also applies when the transient inside the generator fades away. The system-wide control, such as an automatic generation control (AGC), has a much larger time constant than rotor transients, so after the generator transients fades away, the generator models can be converted to PV buses and the following QSS model \cite{dobson2008pserc}, considering the AGC actions, can be used:
	\begin{equation}\label{eqn:eqn_qss}
	\begin{aligned}
	&(P_{Gi}-K_i\Delta f-jQ_{Gi})W^*_i-I_{Li}-\sum_{l}Y_{il}V_l=0\\
	&\dot{P}_{Gi}=-\frac{\Delta f}{T_{gi}}
	\end{aligned}
	\end{equation}
	where $\Delta f$ is the difference between the frequency and the nominal frequency $\Delta f=f-f_s$, $K_i=D_i+1/R_i$ is the coefficient representing the QSS frequency response \cite{ju2018simulation}. $T_{gi}$ is a control time constant of the AGC \cite{dobson2008pserc}.
	
	\subsection{Switching from steady-state to dynamic models}
	When there are no significant fast transients, the QSS simulation can provide satisfactory accuracy and significantly accelerate the computation. When the simulation comes across switch events, the sudden changes in algebraic variables triggers the transient process and the QSS model is converted back to full-dynamic model. The PV buses will be converted to dynamic models of synchronous generators with controllers. 

	\subsection{Efficient determination of steady state using HE coefficients}
	\label{subsec:he_switching}
	\rcolor{Traditional dynamic simulation usually uses the fluctuation of the trajectories to determine the steady state, but it requires an extra period of simulation and is time-consuming. In contrast, HE can enhance the switching by making use of the analytical form of the solutions. Here we propose criteria for determining steady state by using HE coefficients in power series (PS) and Pad\'e appriximation (PA).}
	
	\rcolor{The determination of steady state through PS or PA will need efficient estimation of upper and lower bounds of polynomials within a given interval. So we first provide a general algorithm of estimating such bounds before introducing the steps for determining the steady state. Considering a polynomial $x(t)=\sum_{k=0}^{N}x_kt^k$ and an interval of $t$ as $[0,T]$. First, the polynomial can be written as:}
	\rcolor{\begin{equation}\label{eqn:polynomial_reformat}
	x(t)=(\cdots((x_Nt+x_{N-1})t+x_{N-2})t+\cdots)t+x_0
	\end{equation}}
	\rcolor{here $t$ represents an interval $[0,T]$, and following the interval arithmetic, we can derive the interval of $x_Nt+x_{N-1}$, and then the interval of $(x_Nt+x_{N-1})t+x_{N-2}$, and all the way to the interval of the entire polynomial by unwrapping the parentheses. Then the terminal values of the polynomial are lower and upper bounds of the polynomial. The detailed computation procedures are in Algorithm 1. For a vector of polynomials $\mathbf{y}(t)$ with size $N_y$ and order $N$, the Algorithm 1 has complexity of $O(N_yN)$, which is very efficient.}
	\begin{table}[ht]
		\centering
		\label{tab:alg1}		
		\rcolor{\begin{tabularx}{0.95\linewidth}{X}
			\toprule[1.5pt]
			\textbf{Algorithm 1.} Calculate bounds of polynomial values in given interval. \\ \midrule[1pt]
			\textbf{Input:} Polynomial $x(t)=\sum_{k=0}^{N}x_kt^k$, interval of $t$ as $[0,T]$.\\
			\textbf{Output:} Upper and lower bounds $x_{ub}$, $x_{lb}$, s.t. $x(t)\in [x_{lb},x_{ub}]$ when $t\in[0,T]$.\\ \midrule[1pt]
			\verb| 1| $x_{ub}\gets x_N$, $x_{lb}\gets x_N$\\
			\verb| 2| \textbf{for} $k=N-1 \to 0$ \textbf{do}\\
			\verb| 3| \x \textbf{if} $x_{ub}<0$~~~~~~~~~~~~~~~//Interval arithmetic for $x_{ub}$\\
			\verb| 4| \x\x $x_{ub}\gets x_k$\\
			\verb| 5| \x \textbf{else}\\
			\verb| 6| \x\x $x_{ub}\gets x_{ub}T+x_k$\\
			\verb| 7| \x \textbf{endif}\\
			\verb| 8| \x \textbf{if} $x_{lb}>0$~~~~~~~~~~~~~~~//Interval arithmetic for $x_{lb}$\\
			\verb| 9| \x\x $x_{lb}\gets x_k$\\
			\verb|10| \x \textbf{else}\\
			\verb|11| \x\x $x_{lb}\gets x_{lb}T+x_k$\\
			\verb|12| \x \textbf{endif}\\
			\verb|13| \textbf{end for}\\
			\bottomrule[1.5pt]
		\end{tabularx}}
	\end{table}
	
	\rcolor{Next we introduce the approach for determining steady state using HE coefficients. Assume the power-series approximate solution of a trajectory derived by HE is}
	\rcolor{\begin{equation}\label{eqn:approx_solution}
	x_{T,PS}(t)= \sum_{k=0}^{N}x[k]t^k
	\end{equation}}
	\rcolor{and the solution is effective within the interval $t\in [0,T_e]$, we aim at estimating the rate of changes of $x_{T,PS}(t)$ in $[0,T_e]$. The average rate of change of $x_{T,PS}(t)$ from 0 to $t$ is}	
	\rcolor{\begin{equation}\label{eqn:roc_solution}
	R_{T,PS}(t)\xxdef\frac{x_{T,PS}(t)-x_{T,PS}(0)}{t}=\sum_{k=1}^{N}x[k]t^{k-1}
	\end{equation}}
	\rcolor{So we can use the Algorithm 1 to estimate the bounds of $R_{T,PS}(t)$. Assume the upper and lower bounds of $R_{T,PS}(t)$ are $R_{T,PS,ub}$ and $R_{T,PS,lb}$, respectively. So the average rate of change of $x_{T,PS}(t)$ has a bound $\Delta_{T,PS}=\max\{|R_{T,PS,ub}|,|R_{T,PS,lb}|\}$. If $\Delta_{T,PS}$ is smaller than a preset threshold $\varepsilon_{T}$, then this variable can be considered as entered steady state.}
	
	\rcolor{Besides power series, HE-based simulation usually uses Pad\'e approximations to obtain larger effective time steps than power series. Because power series has smaller effective range, the criteria based on (\ref{eqn:roc_solution}) may be too conservative. So here we propose another criterion for determining the steady state by using the coefficients in Pad\'e approximations. Assume the trajectory of a variable approximated by the Pad\'e approximation is}
	\rcolor{\begin{equation}\label{eqn:approx_solution_pa}
	x_{T,PA}(t)= \frac{\sum_{k=0}^{N_A}x_A[k]t^k}{\sum_{k=0}^{N_B}x_B[k]t^k}
	\end{equation}}
	\rcolor{where $x_A[k]$ and $x_B[k]$ are the Pad\'e coefficients on the numerator and denominator, respectively. To make the Pad\'e approximation unique, it is usually set $x_B[0]=1$. Assume the solution (\ref{eqn:approx_solution_pa}) is effective when $t\leq T_e$. The bounds of (\ref{eqn:approx_solution_pa}) when $0\leq t\leq T_e$ by using the coefficients $x_A[k]$ and $x_B[k]$ will be derived next.}
	
	\rcolor{First we assume the denominator of (\ref{eqn:approx_solution_pa}) $\sum_{k=0}^{N_B}x_B[k]t^k$ does not change sign in interval $[0,T_e]$. This assumption usually holds because if the denominator changes sign, at least one time point in the interval will make the denominator zero, which makes (\ref{eqn:approx_solution_pa}) not well defined for the simulation. So without losing generality, we assume $\sum_{k=0}^{N_B}x_B[k]t^k$ be positive in the interval.}
	
	\rcolor{Second we assume $N_A=N_B$ for the following derivations. If $N_A\neq N_B$ in (\ref{eqn:approx_solution_pa}), e.g. $N_A>N_B$, simply replacing the denominator with $\sum_{k=0}^{N_B'}x_B'[k]t^k$, where $N_B'=N_A$, $x_B'[k]=x_B[k]$ if $k\leq N_B$ and otherwise $x_B'[k]=0$. Denote $c=x_A[0]/x_B[0]$, then (\ref{eqn:approx_solution_pa}) can be written as:}
	\rcolor{\begin{equation}\label{eqn:approx_solution_fluc}
	\resizebox{0.95\hsize}{!}{$\displaystyle{x_{T,PA}(t)=c+\frac{\sum_{k=0}^{N_A}(x_A[k]-cx_B[k])t^k}{\sum_{k=0}^{N_B}x_B[k]t^k}=c+\frac{\sum_{k=1}^{N_A}\tilde{x}_A[k]t^{k-1}}{\sum_{k=0}^{N_B}x_B[k]t^k}t}$}
	\end{equation}	
	where $\tilde{x}_A=x_A[k]-cx_B[k]$. Like (\ref{eqn:roc_solution}), the average rate of change of $x_{T,PA}(t)$ from 0 to $t$ is
	\begin{equation}\label{eqn:roc_solution_pa}
	R_{T,PA}(t)\xxdef\frac{x_{T,PA}(t)-x_{T,PA}(0)}{t}=\frac{\sum_{k=1}^{N_A}\tilde{x}_A[k]t^{k-1}}{\sum_{k=0}^{N_B}x_B[k]t^k}
	\end{equation}}
	
	\rcolor{By using the Algorithm 1, bounds of numerator and denominator in $R_{T,PA}(t)$ can be obtained. Assume the lower bound of $\sum_{k=0}^{N_B}x_B[k]t^k$ is $x_{Blb}$. If $x_{Blb}>0$, i.e. a positive lower bound for the denominator of $R_{T,PA}(t)$ is gotten, then $x_{Blb}$ can be used to estimate the bounds of $R_{T,PA}(t)$. Assume the upper and lower bounds of  $\sum_{k=1}^{N_A}\tilde{x}_A[k]t^{k-1}$ by using Algorithm 1 are $\tilde{x}_{Aub}$ and $\tilde{x}_{Alb}$, respectively, then the upper and lower bounds of $R_{T,PA}(t)$ can be obtained:	
	\begin{equation}\label{eqn:roc_solution_pa_bound}
	\begin{aligned}
	R_{T,PA,lb}&=\frac{\tilde{x}_{Alb}}{x_{Blb}}\\
	R_{T,PA,ub}&=\frac{\tilde{x}_{Aub}}{x_{Blb}}
	\end{aligned}
	\end{equation}
	and the average rate of change of $x_{T,PA}(t)$ is bounded by $\Delta x_{T,PA}=\max\{|R_{T,PA,lb}|,|R_{T,PA,ub}|\}$. And once $\Delta x_{T,PA}<\varepsilon_T$, the studied variable is considered as entered steady state.}
	
	\rcolor{Here we can use both $\Delta x_{T,PA}$ and  $\Delta x_{T,PS}$ for determining the steady state of variables. For a given variable, if $\Delta x_{T,PA}<\varepsilon_T$ or $\Delta x_{T,PS}<\varepsilon_T$, the variable is considered as entered steady state. Note that if $x_{Blb}\leq0$, $\Delta x_{T,PA}$ is not well defined and thus the Pad\'e approximation cannot be used to estimate the bounds for the studied variable.}
	
	\rcolor{In simulation, $\Delta x_{T,PA}$ and $\Delta x_{T,PS}$ of multiple variables need to be calculated and tracked, and only when all the variables satisfy the criteria above, the system can be considered to be in steady state. Most variables in the computation can directly leverage the above criteria for judging steady state, except the generator rotor angles. Because the center of inertia (COI) of the system may not be rotating at the nominal radius speed, the COI will not stop rotating in the nominal frequency coordinate even if system transients have well damped. Consequently, the rotor angles will keep changing even in the steady state. Therefore, the angle of one rotor will be selected as a reference, and the relative angles will be used in the above criteria.}
	
	\section{Overall Extended-term Simulation Framework}
	\subsection{Event-driven simulation based on HE}
	Generally, there are two categories of events in the extended-term simulation:
	
	1) \textit{System events} represent all the actual events in the system, such as switch actions and ramping start/stop.
	
	2) \textit{Simulation events} correspond to changes in the simulation processes but are not actual events in the system, such as the switching to a new simulation stage and switching between dynamic and steady-state models.
	
\begin{figure}[htb]
		\centering
		\includegraphics[clip=true,scale=0.09]{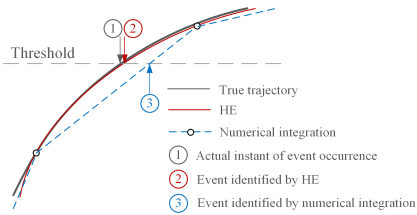}\\
		\caption{Illustration of event-tracking errors using HE and traditional numerical integration simulation methods.}
		\label{fig:he_event}
	\end{figure}

	Some system events are triggered by satisfying some condition $h(\mathbf{x}(t),\mathbf{y}(t),\mathbf{p}(t))\geq 0, t\geq t_0$.
	Once the HE solution of (\ref{eqn:he_sol}) is obtained, the value of $h(\mathbf{x}(t),\mathbf{y}(t),\mathbf{p}(t))$ is tracked by substituting time $t$, and the time of the event can be approximately determined by binary searches. As Fig. \ref{fig:he_event} shows, because HE provides a continuous trajectory of system states in the time domain, HE provides the instant of the event more accurately than the traditional numerical integration methods that only provide values on discrete time steps.
	The simulator uses an event scheduler to manage the event-driven simulation. The event scheduler tracks all the events and determines the instant of the earliest event. At the instant of an event, the simulator simulates the event and the event scheduler updates the event list and prepares for the next event.
	
	\subsection{Overall work flow of extended-term simulation}
\begin{figure}[htb]
		\centering
		\includegraphics[clip=true,scale=0.13]{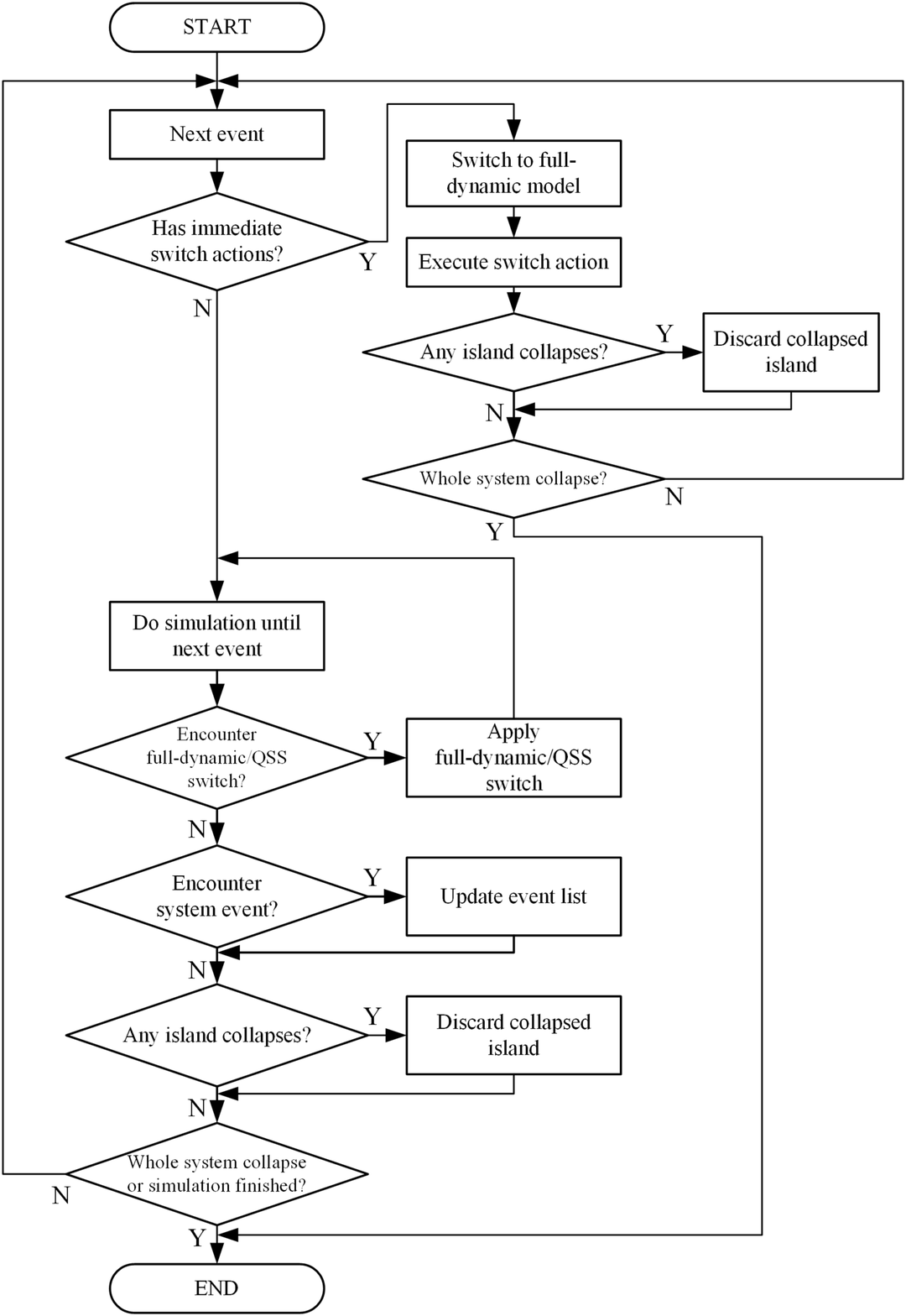}\\
		\caption{Flowchart of extended-term simulation.}
		\label{fig:he_simulation_process}
	\end{figure}

	Fig. \ref{fig:he_simulation_process} shows the overall work flow of the extended-term simulation. The simulation is driven by the event scheduler. The simulation is able to deal with multiple islands and the collapse of part or all of the system.
	
	\section{Test Cases}
	\subsection{2-bus test system}
\begin{figure}[htb]
		\centering
		\includegraphics[clip=true,scale=0.09]{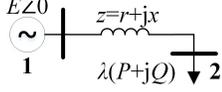}\\
		\caption{2-bus test system.}
		\label{fig:2_bus_system}
	\end{figure}

	We used the 2-bus test system \cite{yao2019efficient} to test the event-driven simulation, as shown in Fig. \ref{fig:2_bus_system}. We use the same system parameters as \cite{yao2019efficient}, i.e. $E=1.01$, $z=r+jx=0.01+j0.05$, and $P+jQ=0.1+j0.3$. Increase the load level $\lambda$ at a constant rate $\diff{\lambda}{t}=1$ from the initial value $\lambda(0)=0$ until voltage collapse. The task is to determine the instant that the current of the line reaches a threshold $I_{th}$. The 2-bus system has a closed-form solution: The square of the line current is
	\begin{equation}\label{eqn:2_bus_current}
	\begin{aligned}
	I^2(t)=&\frac{1}{r^2+x^2}\left[\frac{E^2}{2}-(Pr+Qx)t-\right.\\
	&\left.E\sqrt{\frac{E^2}{4}-(Pr+Qx)t-\frac{(Qr-Px)^2}{E^2}}t^2\right]
	\end{aligned}
	\end{equation}
	and by solving $I^2(t)=I_{th}^2$, the instant of the event $t_{th}$ is
	\begin{equation}\label{eqn:2_bus_time}
	t_{th}=\frac{\sqrt{b^2-4ac}-b}{2a}
	\end{equation}
	where $a=(Pr+Qx)^2+(Qr-Px)^2$, $b=2(Pr+Qx)(r^2+x^2)I_{th}^2$, $c=(r^2+x^2)^2I_{th}^4-E^2(r^2+x^2)I_{th}^2$.
	
\begin{figure}[htb]
		\centering
		\includegraphics[clip=true,scale=0.055]{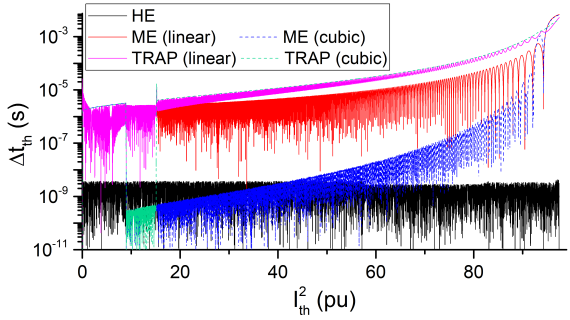}\\
		\caption{\rcolor{Event detection time error $\Delta t_{th}$ with HE, modified Euler (ME) and trapezoidal (TRAP) methods (with linear and cubic interpolation).}}
		\label{fig:2_bus_dt}
	\end{figure}

	\rcolor{Because HE provides the trajectory of states as a continuous function of time, the time of an event can be determined at arbitrarily high resolution with binary search. The modified Euler (ME) and trapezoidal (TRAP) methods with time step 0.01 s are used for comparison. In numerical integration methods, when the threshold falls between two adjacent time steps, the time of an event needs to be approximately determined by interpolation. With the ground-truth solution (\ref{eqn:2_bus_time}), we can compare the error of event time $\Delta t_{th}$ as determined by different methods. The results in Fig. \ref{fig:2_bus_dt} indicate that the traditional methods may have substantial error of $t_{th}$, while HE has very stable and high accuracy for determining the time of event occurrence.}
	
	This test case verifies the reliable performance of HE method over the traditional numerical integration approaches in event-driven simulation. In the next sections, the extended-term hybrid steady-state and dynamic simulation approach will be demonstrated. As we have shown the reliability and the superior performance of HE-based simulation approach, we will compare the hybrid steady-state and dynamic simulation approach with the full-dynamic HE simulation approach in the next sections. 
	
	\subsection{4-bus test system}
\begin{figure}[htb]
		\centering
		\includegraphics[clip=true,scale=0.07]{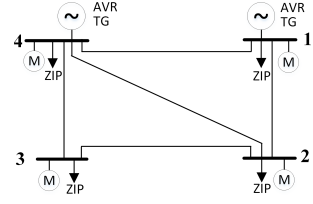}\\
		\caption{4-bus test system.}
		\label{fig:sys_4}
	\end{figure}

	In this subsection, we show the proposed extended-term simulation method on a 4-bus test system. As Fig. \ref{fig:sys_4} shows, each bus has ZIP and induction motor loads, and buses 1 and 4 have synchronous generators with AVRs and TGs. The TGs have time constants $T_1=0.3$ s, $T_2=0.1$ s. The system is also equipped with AGC, and the time constant of the AGC controller is $T_{gi}=5$ s. In the beginning, each generator has an active power output of 1.1436 pu.  
	
	During the 500 s simulation, the system periodically adds and cuts loads on buses 2 and 3 at time intervals of 30 s. The process involves multiple events, fast transients of generators and induction motors, and slower dynamics introduced by AGC, and the duration is much longer than the conventional dynamic security assessment. People can choose to perform the HE full-dynamic simulation with the conventional numerical simulation approaches or, alternatively, HE can be used to perform simulation for better accuracy and computation speed \cite{yao2019efficient}. However, the simulation is still time-consuming for such a long process. Here we use the proposed approach for switching between dynamic and QSS models based on HE, and compare it with the HE full-dynamic simulation results. 
	\begin{figure}[htb]
		\centering
		\includegraphics[clip=true,scale=0.08]{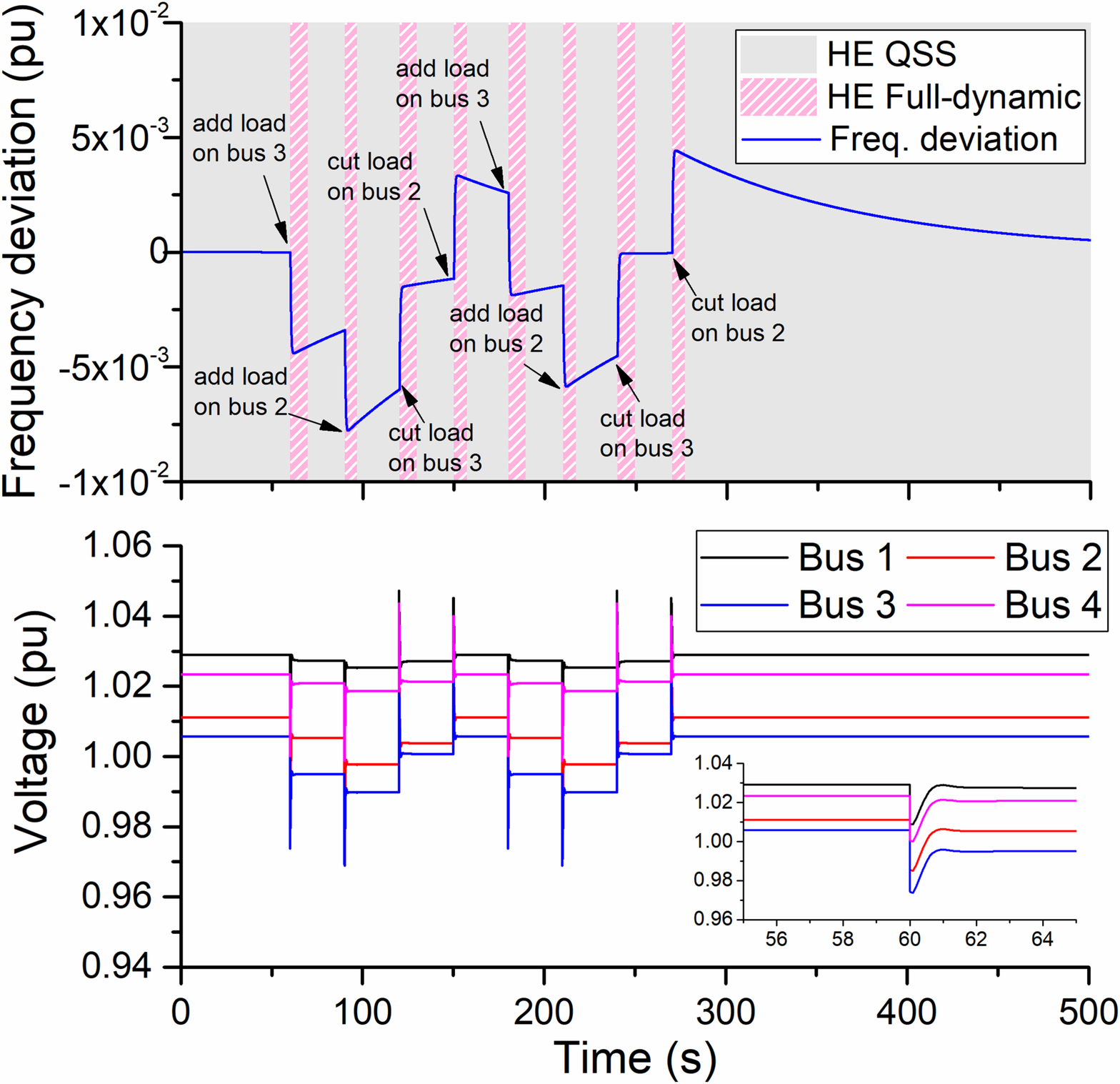}\\
		\caption{Frequency and voltage of 4-bus system.}
		\label{fig:4_bus_freq_volt}
	\end{figure} 
	\begin{figure}[htb]
		\centering
		\includegraphics[clip=true,scale=0.07]{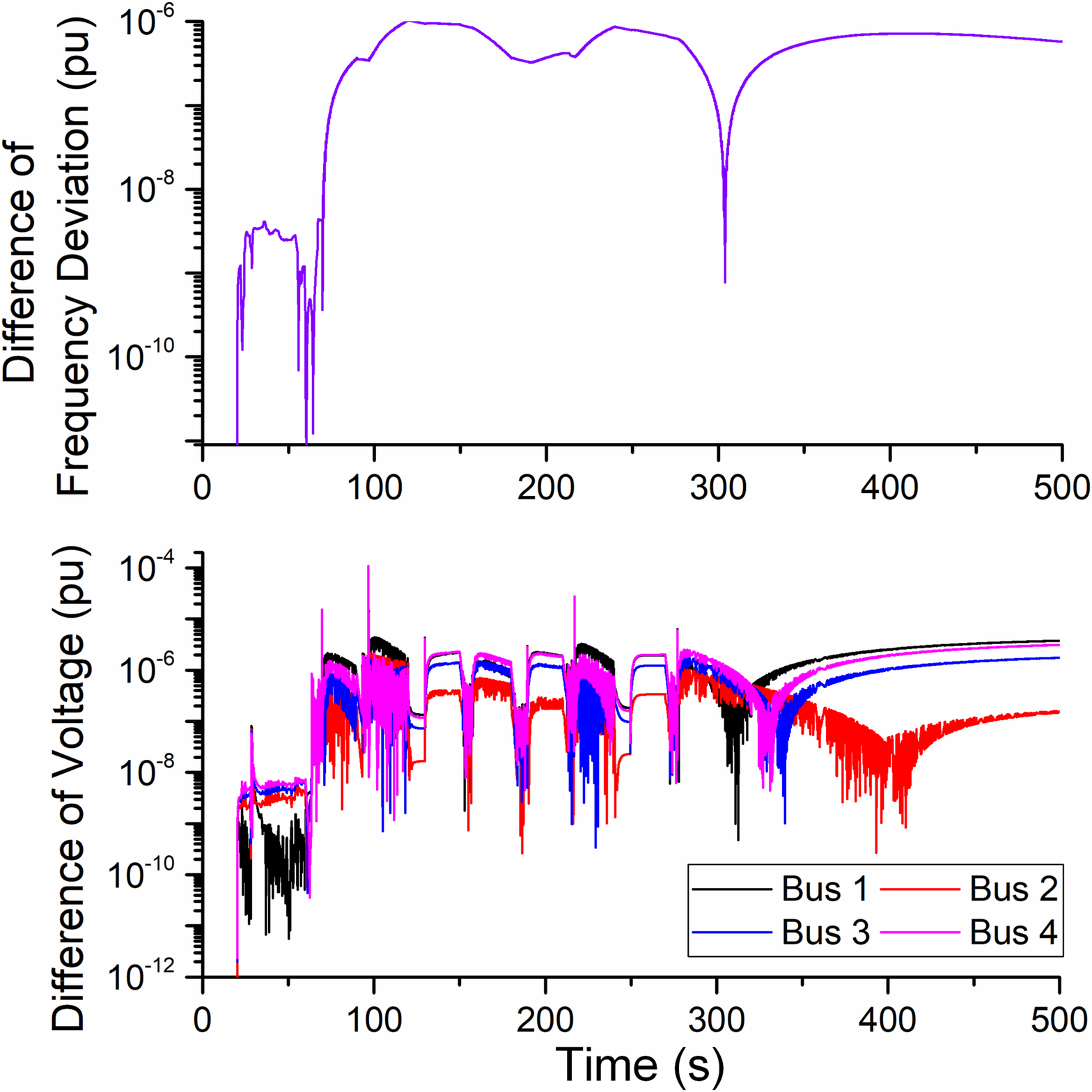}\\
		\caption{\rcolor{Difference of frequency and voltage between HE full dynamic and hybrid simulation on 4-bus system.}}
		\label{fig:4_bus_diff_freq_volt}
	\end{figure} 
	
	\rcolor{Fig. \ref{fig:4_bus_freq_volt} shows the frequency and voltage of the system (for full dynamic simulation, the frequency is regarded as the mean value of the generator rotor speeds weighted by the inertia). The frequency curve clearly reflects the tendency of AGC to restore system frequency. The figure also shows the time intervals of the full-dynamic simulation and QSS simulation. In the whole 500 s process, only 66.32 s are in full-dynamic simulation, and about 87.7\% of the process is simulated with QSS. The HE QSS+full-dynamic hybrid simulation is significantly faster than the HE full-dynamic simulation: Full-dynamic simulation takes 47.32 s, and hybrid simulation only takes 12.66 s, which reduces the computation time by 73.25\%. Fig. \ref{fig:4_bus_diff_freq_volt} shows the difference of the simulation results between full-dynamic simulation and hybrid simulation: The result of hybrid simulation is almost the same with that of the full-dynamic simulation, which verifies that hybrid simulation can well reproduce the result of full-dynamic simulation. }
	
	\rcolor{To better explain the HE-based switching from dynamic to QSS models, some HE coefficients at the model switching point of 69.69 s are extracted. At this time point, the simulation switches from dynamic model to the QSS model. The HE solution has an effective range $T_e=0.137$ s. Here the difference of rotor speeds between the two generators $\omega'$, the square of voltage magnitude at bus 4 $V^2_4$ and AVR variable on generator 2 $v_{m2}$ (on bus 4) are listed in Table \ref{tab:he_coeff_4}. The threshold for determining steady state is selected as $\varepsilon_T=10^{-3}$. The results show that for $\omega'$, both $\Delta x_{T,PS}$ and $\Delta x_{T,PA}$ satisfy the steady-state criteria, while only $\Delta x_{T,PA}$ for $V^2_4$ and $\Delta x_{T,PS}$ for $v_{m2}$ satisfy the steady-state criteria. In any those cases, the variable are determined as entered steady state because at least one from $\Delta x_{T,PS}$ and $\Delta x_{T,PA}$ satisfy the criteria. This shows that both $\Delta x_{T,PS}$ and $\Delta x_{T,PA}$ are useful for the effective switching from dynamic to QSS simulation. }
	
	\begin{table}[h]
		\centering
		\caption{HE coefficients (power series) of rotor speed difference in 4-bus system}
		\label{tab:he_coeff_4}
		\small
		\rcolor{\begin{tabularx}{0.8\linewidth}{p{1.3cm}@{}*3{>{\centering\arraybackslash}X}@{}}
			\toprule[1pt]
			$k$               & $\omega'$        & $V^2_4$          & $v_{m2}$      \\\midrule[0.5pt]
			$\Delta x_{T,PS}$ & \sci{6.11}{-4}   & 0.0279           & \sci{3.76}{-4}\\
			$\Delta x_{T,PA}$ & \sci{1.26}{-4}   & \sci{9.85}{-4}   & 0.0013        \\
			\bottomrule[1pt]
		\end{tabularx}}
	\end{table} 

	\subsection{Simulation of restoration on New England test system}
	Finally, we simulated and demonstrated the process of system restoration on the IEEE 39-bus (New England) system. System restoration is a typical process involving complex dynamics in different timescales, with significant topology and system parameter changes, which is very challenging to simulate \cite{qiu2017integrated,hou2011computation}. The generator on bus 39 acts as the black start generator for the restoration process. The buses, lines and generators are energized sequentially, and the loads and generation are picked up gradually. In this simulation task, the generators use 6th-order model, and the loads use ZIP+Motor model. The entire restoration process lasts 12,065 s, and full-dynamic and hybrid simulation approaches are used to simulate them respectively. The simulation was implemented and tested on Matlab 2017b.
	\begin{figure}[htb]
		\centering
		\includegraphics[clip=true,scale=0.1]{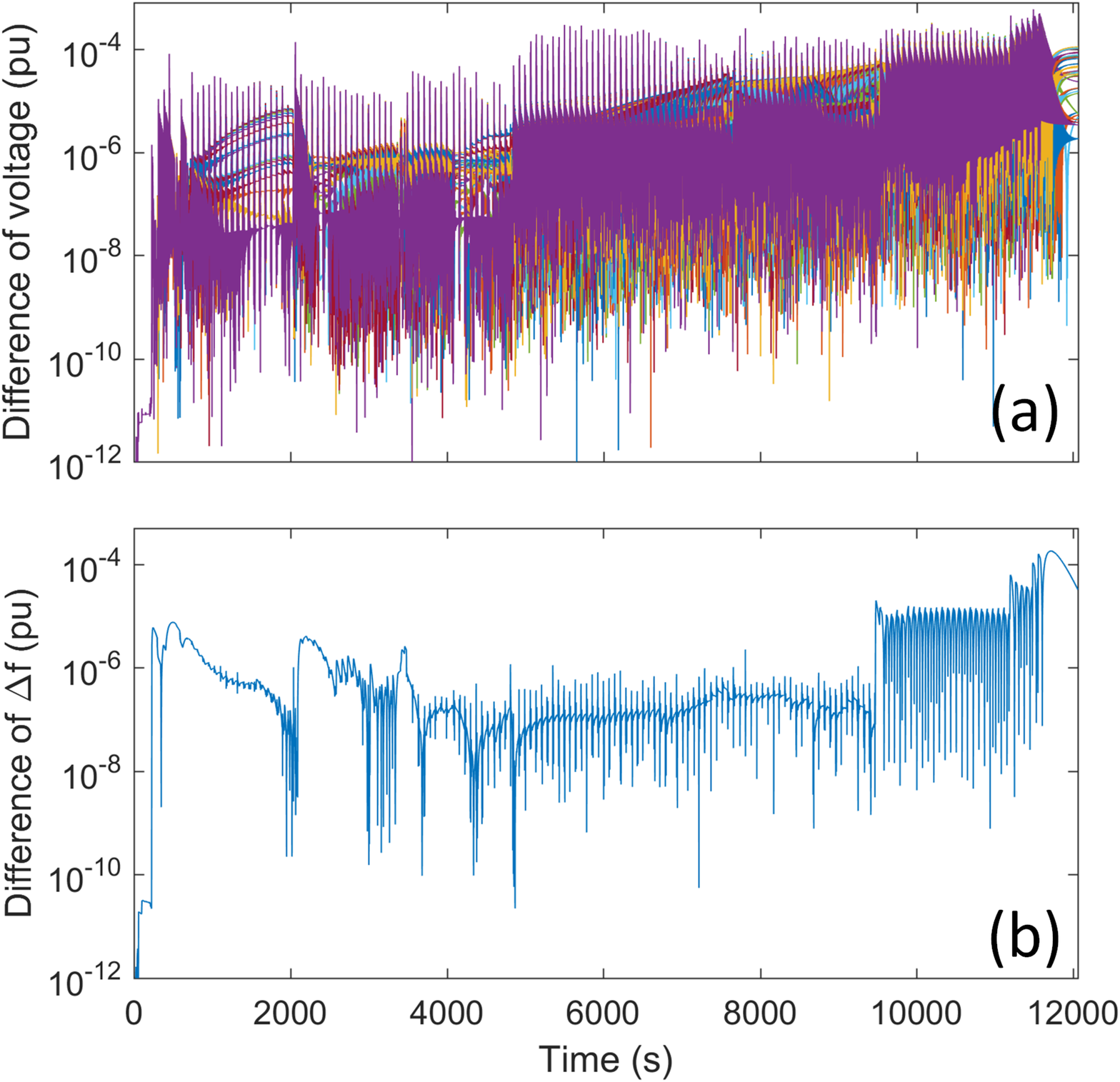}\\
		\caption{Difference of (a) voltage and (b) frequency between HE full-dynamic and HE hybrid simulation on 39-bus system.}
		\label{fig:039_restoration_deltavf}
	\end{figure}
	\begin{figure*}[h]
		\centering
		\includegraphics[width=0.8\textwidth]{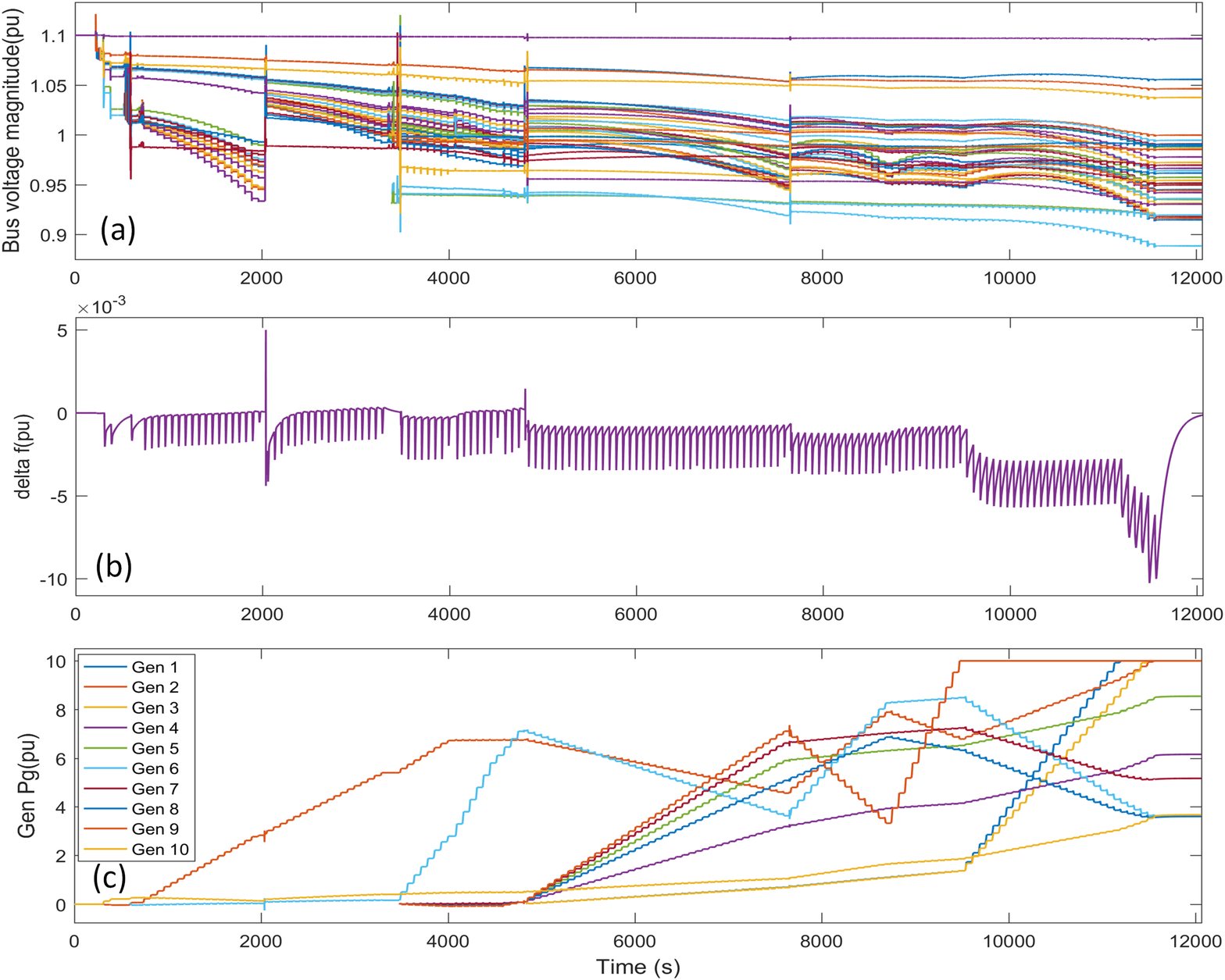}
		\caption{Selected system state trajectory of IEEE 39-bus system under restoration: (a) Voltage; (b) Frequency; (c) Reference generator active power.}
		\label{fig:039_restoration_vfpg}
	\end{figure*}
	During the whole 12,065 s restoration process there are 396 events, including adding lines, generators, static load, shunt capacitors, and induction motors, and ramping up generator power. In the hybrid simulation, 8,057.3 s (i.e., 66.8\%) of the restoration process is simulated in a QSS model. Fig. \ref{fig:039_restoration_deltavf} shows the difference in voltage and frequency between the full-dynamic simulation and the hybrid simulation, and the trajectories of some system states are shown in Fig. \ref{fig:039_restoration_vfpg}.  It can be seen that the results of hybrid simulation are very close to those of full-dynamic simulation. In terms of the computation speed, the full-dynamic simulation takes 5,909.3 s to finish computation. The hybrid simulation takes 2,779.2 s, which is a time savings of about 52.97\%. Considering that QSS simulation covers 66.8\% of the entire restoration process, the result shows that QSS simulation is significantly faster than full-dynamic simulation, and the hybrid simulation approach can significantly  enhance the performance of extended-term simulation without losing accuracy.
	
	\section{Conclusion}
	This paper proposes a novel extended-term simulation approach based on holomorphic embedding (HE). The high accuracy and efficiency of HE lays the foundation for the extended-term simulation. The HE formulations for solving typical types of events (i.e., atomic events) are provided. And to efficiently accelerate simulation under multiple timescales, the hybrid simulation of dynamic and quasi-steady-state (QSS) simulation based on HE is proposed, and its accuracy has been verified. The model switching between dynamic and QSS simulation based on HE coefficients is efficient and convenient. The nature of the HE solution as continuous functions in a time domain also enables better handling of various events in the extended-term simulation. The test cases show that the proposed hybrid, event-driven, extended-term simulation based on HE has satisfactory accuracy and efficiency and can be used to simulate complex power system processes such as restoration, cascading outages and renewable energy control.

	
	%

	


	\ifCLASSOPTIONcaptionsoff
	\newpage
	\fi

	
	
	%
	
	\bibliographystyle{IEEEtran}
	\bibliography{bib/refs}
	
	%
	
	
	
	
	
	
	

\end{document}